\documentclass[prl,aps,showpacs,superscriptaddress,twocolumn]{revtex4}
\usepackage{color} 
\usepackage{bm}
\usepackage{dcolumn}
\usepackage{times}
\usepackage{amssymb} 
\usepackage{amsmath} 
\usepackage{ragged2e}
\usepackage{graphicx}
\usepackage{epstopdf}
\usepackage[english]{babel}
\usepackage{mathrsfs}
\usepackage{textcomp}
\usepackage{amsthm}
\usepackage{amsfonts}
\usepackage{soul}
\usepackage{braket} 
\usepackage{hyperref}
\usepackage{cleveref}
\newcommand{\be}{\begin{equation}}
\newcommand{\ee}{\end{equation}}
\newcommand{\bea}{\begin{eqnarray}}
\newcommand{\eea}{\end{eqnarray}}
\newcommand{\ba}{\begin{array}}
\newcommand{\ea}{\end{array}}

\begin{document}

\title{Interacting second-order topological insulators in one-dimensional fermions \\with correlated hopping}
\author{A. Montorsi}
\affiliation{Institute for Condensed Matter Physics and Complex Systems, DISAT, Politecnico di Torino, I-10129 Torino, Italy}

\author{U. Bhattacharya}
\affiliation{ICFO - Institut de Ci\`encies Fot\`oniques, The Barcelona Institute of Science and Technology, 08860 Castelldefels (Barcelona), Spain.}

\author{Daniel González-Cuadra}
\affiliation{Institute for Theoretical Physics, University of Innsbruck, 6020 Innsbruck, Austria}
\affiliation{Institute for Quantum Optics and Quantum Information of the Austrian Academy of Sciences, 6020 Innsbruck, Austria}

\author{M. Lewenstein}
\affiliation{ICFO - Institut de Ci\`encies Fot\`oniques, The Barcelona Institute of Science and Technology, 08860 Castelldefels (Barcelona), Spain.}
\affiliation{ICREA, Pg. Lluis Companys 23, ES-08010 Barcelona, Spain.}

\author{G. Palumbo}
\affiliation{School of Theoretical Physics, Dublin Institute for Advanced Studies, 10 Burlington Road, Dublin 4, Ireland.}

\author{L. Barbiero}
\affiliation{Institute for Condensed Matter Physics and Complex Systems, DISAT, Politecnico di Torino, I-10129 Torino, Italy}

\begin{abstract}
\noindent Higher-order topological crystalline phases in low-dimensional interacting quantum systems represent a challenging and largely unexplored research topic. Here, we derive a Hamiltonian describing fermions interacting through correlated hopping processes that break chiral invariance, but preserve both inversion and time-reversal symmetries.
In this way, we show that our one-dimensional model gives rise to an interacting second-order topological insulating phase that supports gapped edge states.
The topological nature of such interacting phase turns out to be revealed by both long-range order of a non-local string correlation function and by even degeneracy of the entanglement spectrum. For strong interactions we instead find that the topological crystalline phase is destroyed and replaced by a singlet superconducting phase. The latter, characterized by local fermionic pairing, turns out to appear both in a homogeneous and in a phase separated form. Relevantly, the derived one-dimensional model and the second-order topological insulator can be explored and investigated in atomic quantum simulators.      
\end{abstract}

\maketitle
\paragraph{\textbf{Introduction}.}
Symmetries represent a fundamental tool to characterize quantum systems \cite{gross1996}. The concept of spontaneous symmetry breaking has indeed allowed to define the large majority of states of matter by means of specific local order parameters \cite{Landau1937}. Escaping from this classification, the recently discovered symmetry protected topological (SPT) states of matter are gaining intense attention \cite{pollmann2010,gu,chiu2016,montorsi}. SPT phases result to be deeply understood at single-particle level, where exact solutions can be derived \cite{kitaev2009,kitaev2,altland}. Specifically to one-dimensional systems, it has been initially demonstrated that chiral symmetry represents a strict requirement in order to achieve robust topological insulating states \cite{kitaev2009}. Such symmetry can formally occur when time-reversal and particle-hole symmetries are simultaneously either fulfilled or absent. Recently, a class of topological gapped phases characterized by crystalline symmetries has revealed further interesting features related to the presence of space-group symmetries \cite{Fu2011,Fang2012,Slager2013, Slager2017, Neupert2018}. 
More precisely, it has been unveiled in non-interacting models that the inversion symmetry can still allow the conservation of topological insulating properties even when particle-hole and chiral invariance are not fulfilled \cite{Neupert2018,Prodan2011, Alex2014, fang2021,cano2022,vanmiert2017,Miert_2016}.
The key role of inversion symmetry has been further explored in high-dimensional systems in the context of fragile topology both in free \cite{Po2018, Cano2018,Bradlyn2019,Bouhon2019,https://doi.org/10.48550/arxiv.1810.02373,Ahn2019, Hwang2019,Wieder2020,Bouhon2020} and in interacting systems \cite{Else2019,Liu2019,Latimer2021,Joshi2021, Peri2021, Turner2022}. Moreover, it has been shown that the combination of inversion and time reversal symmetries plays a central role in various kinds of higher-order topological insulators (HOTIs) in high dimensionality \cite{Schindler2018, Khalaf2018, Wang2019, Lin2021, Lin2022,Benalcazar2017,Langbehn2017,Miert2018,Calugaru2019, DiLiberto2020,Hwang2019}.\\Topological states of matter result to be much less understood in presence of interactions. In the one dimensional case, a systematic classification of possible SPT phases has been given in \cite{montorsi} in the low energy limit. Moreover, numerical methods have revealed a plethora of interacting SPT insulators \cite{haldane,dalla,dalmonte,furukawa,koba,barbiero,fazzini,barbiero1,Gonzalez-Cuadra_2019a,Barbarino2019,Gonzalez-Cuadra_2019b}, superconductors \cite{fazzini2,fazzini3,Hassler_2012,thomale2013,katsura2015,miao2017} and critical regimes \cite{ruben,verresen2021,ruben3,Fraxanet2022}. Nevertheless, low dimensional higher-order topological crystalline phases remain an unexplored playground. Crucially, atomic quantum simulators have allowed to shed further light on this challenging topic by experimentally realize interacting topological matter \cite{semeghini2021,Browaeys2020,sompet2022,walter2022}. Nevertheless, the experimental implementation of both single-particle and interacting higher order topological phases remain an unexplored playground.\\ The impressive accuracy in Hamiltonian engineering and measurement protocols further placed ultracold atomic platforms among the most efficient ways to investigate symmetry breaking quantum states. As prominent examples, specific designs of the interacting processes have allowed to realize phases with broken translational \cite{schauss2012,li2017,leonard2017,guo,tanzi,chomaz}, time-reversal \cite{tai2017,wang2021} and spin rotational \cite{Mazurenko17} symmetry. Moreover, different schemes employed in ultracold atomic quantum simulators make possible to design a peculiar kind of interaction: the correlated hopping (CH) processes also called the bond-charge interaction  \cite{ma2011,Dobry2011,jurgensen2014,meinert2016,ferlaino,gorg2018,messner2018}. Specifically to fermionic systems, CH results to be of great relevance as it breaks the particle-hole (p-h) symmetry but preserves the time-reversal invariance. The role of such interaction has been originally investigated in the context of hole-superconductivity \cite{hirsch1989} while, more recently, it has been realized that a large variety of phenomena including Peierls instability \cite{aligia2007}, nanoscale phase separation \cite{anfossi2009}, quantum many-body scar states \cite{mark2020,hudomal2020} and lattice gauge theories \cite{Barbiero2019,Schweizer2019,Gorg2019} can be explored in quantum many-body systems characterized by the presence of CH.\\ 
\noindent Motivated by the fact that higher-order one-dimensional interacting topological crystalline phases with broken chiral symmetry remain both theoretically and experimentally largely unexplored, we derive a new model of ultracold fermions in optical lattice that supports the appearance of an interacting generalization of a second-order topological insulator in one dimension \cite{Hwang2019}.
Notice that in 1D single-particle second-order topological phases, namely obstructed-atomic insulators \cite{Bradlyn2017}, there exists a mismatch between the atomic and Wannier centers of the given unit cell. In the presence of interactions, the concept of location of the Wannier functions is not well defined anymore. Thus, we identify interacting second-order topological insulators through different theoretical tools, such as non-local string correlation and entanglement spectrum.

In our model, we consider fermions tunneling in a dimerized lattice geometry and interacting through CH processes. By means of density-matrix-renormalization-group (DMRG) calculations \cite{dmrg}, we reveal that such model supports the presence of an interacting second-order topological insulator with gapped edge states and broken chiral symmetry. Furthermore, we find that when the CH is strong, the topological properties disappear and a topologically-trivial superconducting phase takes place. The latter results to be characterized by fermionic singlet states, therefore representing an example of Luther-Emery superconductivity \cite{luther1974}. Noticeably, we confirm the presence of such phase transition both by DMRG and by an analytical treatment based on a slave-boson mapping. Our results further point out that a larger strength of the CH allows to achieve a regime of singlet superconductivity associated to an inhomogeneous density distribution as unveiled by the infinite compressibility. In order to complement the equilibrium physics discussed above, we show that the interacting second-order topological insulator remains stable even in out-of-equilibrium schemes where the chiral symmetry is broken by performing a Hamiltonian quench. Finally, we discuss how our results can be probed in the ongoing experimental platforms based on ultracold atoms in optical lattice.

\begin{figure}
\includegraphics[width=\columnwidth]{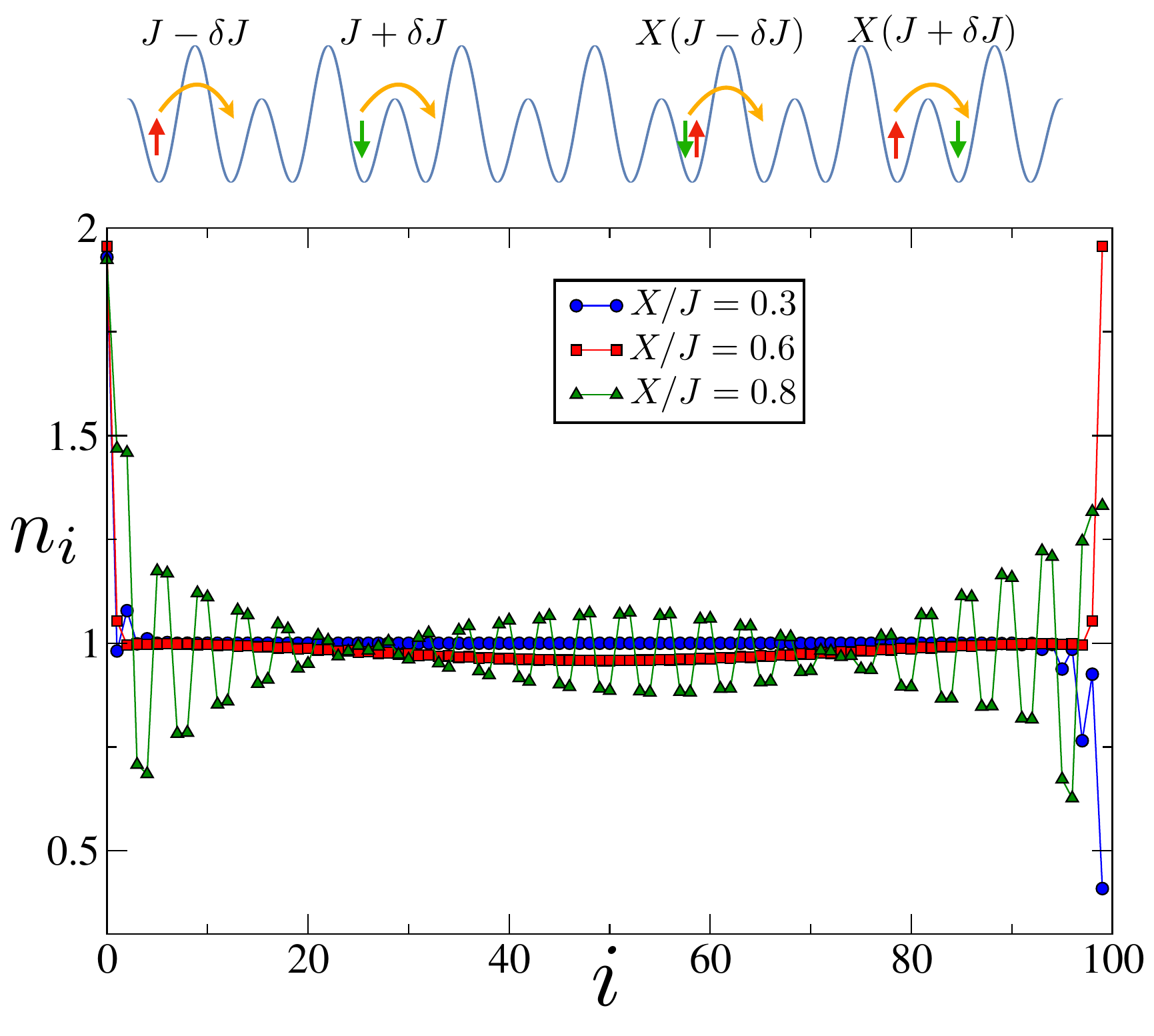}
\caption{(Color online). Cartoon of the model and DMRG density profile $n_i$ obtained by fixing $N=2N_{\uparrow}=2N_{\downarrow}=L=100$, $\delta J/J=0.4$ and different values of $X/J$. In order to break the ground state degeneracy here and in all the other figures we apply chemical potentials of the order of $10^{-2}J$ with opposite sign in the first and last sites.}
\label{fig1}
\end{figure}

\paragraph{\textbf{Model}.}
As depicted in Fig. \ref{fig1}, we investigate a model consisting of $N$ fermions described by the usual creation (annihilation) operators $c^\dagger_{i,\sigma}$ ($c_{i,\sigma}$) where $\sigma=\uparrow,\downarrow$ labels the the two possible fermionic internal states and $i$ refers to the $i$-th site of a dimerized one dimensional lattice of length $L$. Such Hamiltonian reads as
\begin{eqnarray}
H&&=\sum_{i,\sigma}\Big[-\big(J+\delta J(-1)^i\big)\big(c_{i,\sigma}^\dagger c_{i+1,\sigma}+h.c.\big)+\\
\nonumber
&&+X\big(J+\delta J(-1)^i\big)\big(n_{i,\bar{\sigma}}+n_{i+1,\bar{\sigma}}\big)\big(c_{i,\sigma}^\dagger c_{i+1,\sigma}+h.c.\big)\Big]
\label{ssh}
\end{eqnarray}
where we fix the particle density $\bar{n}=N/L=2N_{\uparrow}/L=2N_{\downarrow}/L=1$ with $\sum_in_{i\uparrow(\downarrow)}=N_{\uparrow(\downarrow)}$. The terms in the first line, $J-\delta J$ $(J+\delta J)$ represent the single particle tunneling probability between two lattice sites connected by odd (even) links. The interaction between fermions is instead captured by the term $X(J-\delta J)$ $(X(J+\delta J))$ describing correlated hoppings between two sites connected by an odd (even) link. More precisely, this last term takes into account the system energy variation when a fermion tunnels from/to an already occupied site.\\The model eq. (\ref{ssh}) at $X=0$ is the well known Su-Schrieffer-Heeger (SSH) Hamiltonian \cite{SSH1979}. At half filling $\bar{n}=1$, the latter is chiral symmetric as both p-h and time reversal symmetries are fulfilled. In addition to this, for $\delta J>0$ it can be shown that a first-order topological insulator of the BDI class \cite{altland} protected by chiral symmetry and hosting gapless degenerate topological edge states occur.  It is worth to underline that, as shown in different previous cases \cite{Gurarie2011,Manmana2012,Wang2015,Yoshida2014,Ye2016,yu,Barbiero2018,Sbierski2018,Le2020}, the topological insulating regime of the SSH model remains stable even in presence of interacting terms that preserve the chiral symmetry. However, as previously specified, the correlated hopping processes do not fulfill this last criterion since for $0<|X/J|<1$, particle-hole symmetry and thus chiral symmetry are broken. For this reason, it is relevant to understand whether the Hamiltonian eq. (\ref{ssh}) can still support the presence of protected topological edge states.    

\paragraph{\textbf{Second-order topological crystalline insulator}.}

\begin{figure}
\includegraphics[width=\columnwidth]{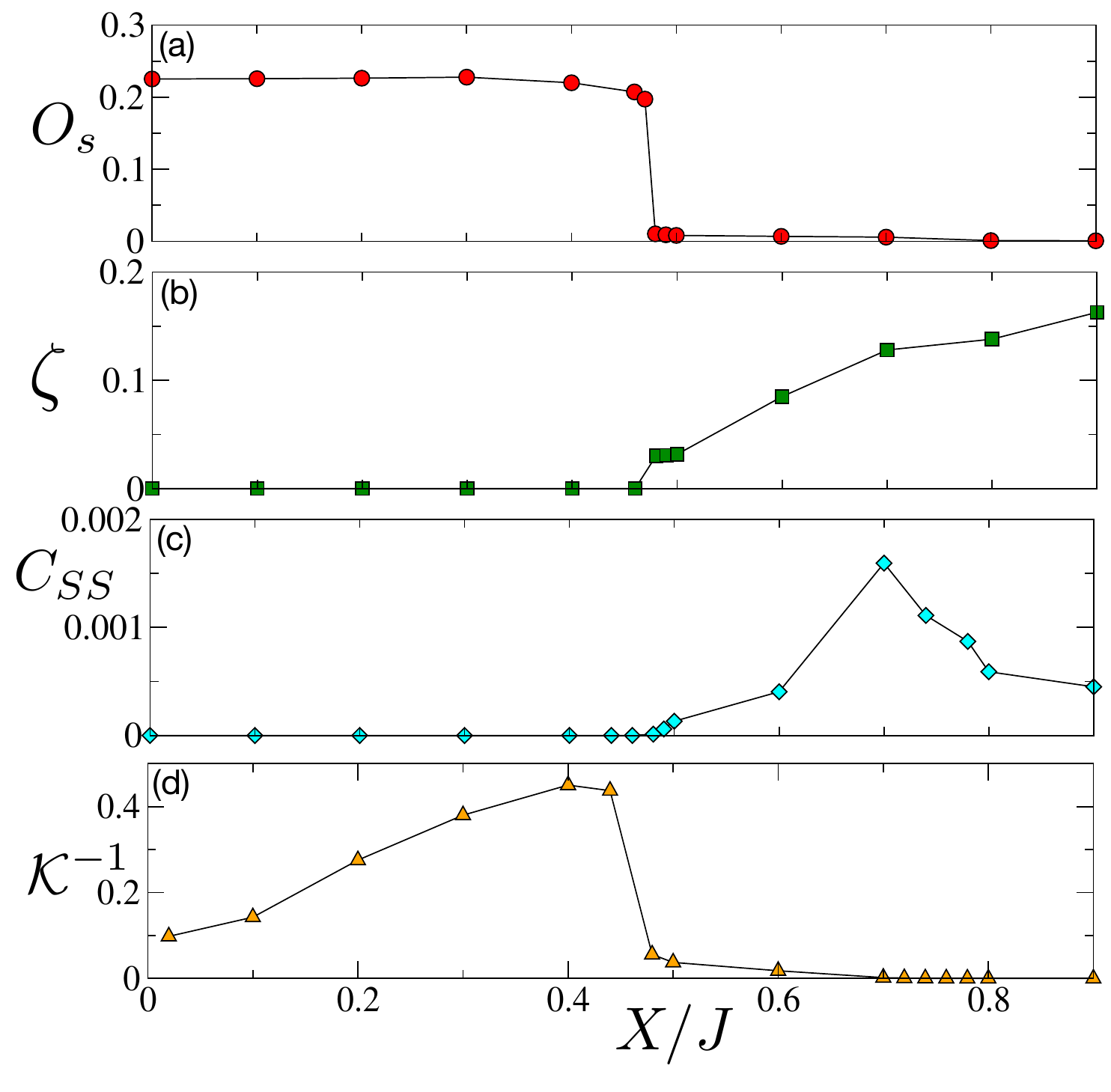}
\caption{(Color online). \textit{a)} String order correlation function eq. (\ref{string}) extracted at $|2i+1-2j|=40$ considering only the 80 central sites. \textit{b)} Entanglement spectrum $\zeta=\lambda_1^N-\lambda_2^N+\lambda_3^N-\lambda_4^N$ extracted by considering a system bipartition equal to $L/2$. \textit{c)} Singlet superconducting correlator eq. (\ref{ss}) extracted at $|i-j|=80$. \textit{d)} Compressibility eq. (\ref{comp}) extracted by considering variations in the number of particles equal to 2 and 4. The results refer to a configuration with $N=2N_{\uparrow}=2N_{\downarrow}=L=100$, $\delta J/J=0.4$ and different values of $X/J$}
\label{fig2}
\end{figure}

We begin our analysis by fixing the fermionic density $\bar{n}=1$ and $\delta J=0.4$ such that for $X=0$ the ground state is a topological insulator protected by chiral symmetry. In Fig. \ref{fig1}, we compute the density profile $n_i=\sum_{\sigma}n_{i,\sigma}$ and notice the presence of three very different regimes. For large $X/J$ a peculiar global and local modulation in the fermionic distribution occurs. This last behavior is lost at intermediate value of $X/J$ where a symmetric particle accumulation takes place at the edges and the density distribution in the bulk is lower than one. Interestingly, for smaller strength of the CH interaction, our results show the presence of antisymmetric fermionic accumulation at the edges of the system. Furthermore, we find that in the bulk of the chain the density profile $n_i=1$, thus being compatible with topologically gapped bulk excitations. As already mentioned, this last scenario characterizes also the $X=0$ case where the system is a first-order topological insulator protected by chiral symmetry. As strict probes to understand whether a topological insulating regime exists even for broken chiral symmetry, i. e. when $X/J\neq0$, in Fig. \ref{fig2} we plot the non-local string correlation function capturing the topological order of the SSH model \cite{Anfuso2007,Wang2015,Fraxanet2022,sergi2022}
\begin{equation}
O_S=\Big(-4\big\langle S^c_{2i+1}\exp[\imath\pi\sum_{k=2i+2}^{2j-1}S^c_k]S^c_{2j}\big\rangle\Big)
\label{string}
\end{equation}
and the entanglement spectrum \cite{li2008,pollmann2010,turner2011,kitaev2}
\begin{equation} 
\rho_A=\sum_{N,n}\lambda_n^N\rho_n^N
\label{es}
\end{equation}
where $\rho_A$ is the density matrix of a system bi-partition $A$ and $\rho_n^N$ describes a pure state of $N$ particles with corresponding eigenvalues $\lambda_n^N$, namely the entanglement spectrum (ES). In particular, long-range order of uniquely eq. \ref{string} reflects the peculiar non-local nature of topological gapped bulk excitations and the even degeneracy of eq. \ref{es} captured by the vanishing value of $\zeta=\lambda_1^N-\lambda_2^N+\lambda_3^N-\lambda_4^N$ witnesses the degeneracy of the topological edge states. As shown in the upper panels of Fig.~\ref{fig2}, the two aforementioned requirements are totally fulfilled and thus unambiguously reveal the existence of a topological insulator not invariant under chiral symmetry, i.e. a second-order topological insulating phase \cite{Hwang2019}. The appearance of such topological phase is explained by the fact that the Hamiltonian in eq. (\ref{ssh}) consists of alternating bonds being centers of inversion thus resulting invariant under inversion symmetry. As a consequence of this crystalline symmetry, our topological phase can be seen as an interacting version of a single-particle obstructed-atomic insulator \cite{Bradlyn2017} in which its corresponding Wannier centers cannot be continuously moved to match with the particle positions while obeying inversion symmetry and without closing the gap. This implies that even in the absence of chiral symmetry, the corresponding topological invariant in the bulk remains quantized.
The results in Fig. \ref{fig2} clarify that such topological crystalline state is robust for $X/J\lesssim 0.5$ and we checked that such value remains stable for different choices of $\delta J>0$. Indeed, by further increasing the CH interaction strength both the long-range order of $O_S$ and the vanishing value of $\zeta$ are lost therefore signaling that a phase transition between a gapped topological and a trivial state has occurred. In order to confirm the disappearance of the topological insulating state, we perform a mapping of the model eq. (\ref{ssh}) following the scheme derived in \cite{roncaglia2010} where the case of $\delta=0$ and finite onsite repulsion has been treated. In particular, by the employing a slave-boson formalism, holding exactly at $X/J=1$, to lower $X/J$, eq. \ref{ssh} can be rewritten in terms of spinless fermions 
\begin{equation}
H_{eff}=-X\sum_{i}\bigl[1+(-)^{i}\delta J\bigr]\left(f_{i}^{\dagger}f_{i+1}+\gamma f_{i}^{\dagger}f_{i+1}^{\dagger}+\mathrm{H.c.}\right)\:,
\end{equation}
where $\gamma=\frac{J-X}{X}$, and $f_i,f_i^\dagger$ are ladder operators which create/destroy spinless fermions out of empty and doubly occupied sites. The spectrum of $H_{eff}$ is obtained by rotating its Fourier transform into
\begin{equation}
\tilde H_{eff}=\pm X\sum_{k\in BZ}\Lambda_{k}^{\pm}\left(\beta_{k}^{\dagger}\beta_{k}-\frac{1}{2}\right),\label{eq:Hdiag}
\end{equation}
with $\beta_{k},\beta_{k}^{\dagger}$ being the ladder operators for the Bogolubov
fermionic quasiparticles, and $\pm\Lambda_{k}^{\pm}$ the dispersion relations for the four bands. Explicitly
\begin{eqnarray}
\Lambda_{k}^{\pm}&&=2\big((1+\gamma^{2})(1+(\delta J)^2)\pm \nonumber\\
&&\pm2 \gamma\delta J+(1-(\delta J)^2)(1-\gamma^{2})\cos{k}\big)^{1/2}\quad. \label{spec}
\end{eqnarray}
Here, for $\gamma=0$ (i.e. $X/J=1$) we recognize the spectrum of the integrable case \cite{Montorsi_2008}. Moreover, the two values $\gamma=1$
and $\delta J=1$ turn out to be critical. Indeed, $H_{eff}$ can be exactly mapped onto a (staggered) anisotropic XY model, which at $\gamma=1$ (i.e. $X/J=0.5$) is known to exhibit a commensurate-incommensurate transition thus in agreement with our DMRG results in Fig. \ref{fig2}. More precisely, such transition reflects the closing of the topological gap and the opening of a spin gap. Our DMRG results show indeed that this trivial phase is characterized by quasi-long-range order of 
\begin{equation}
C_{SS}=\Big\langle c^\dagger_{i,\sigma}c^\dagger_{i,\bar{\sigma}}c_{j,\bar{\sigma}}c_{j,\sigma}\Big\rangle
\label{ss}
\end{equation}
which can be produced uniquely by a finite spin gap. In particular, the above behavior is a well established signature of the appearance of superconducting states where fermions with opposite spin are paired and form bounded singlets. Noticeably, such singlet superconductor phase is analogous to that occurring, for instance, in the attractive Hubbard model and in the more general Luther-Emery liquid phase \cite{luther1974} of the sine-Gordon model \cite{Giamarchi2004}. While such superconducting order persists in the whole region $0.5<X/J<1$, we notice that around $X/J\approx2/3$ a regime characterized by vanishing inverse compressibility 
\begin{equation}
{\cal{K}}^{-1}=\bar{n}^2\frac{\partial^2E_0}{\partial \bar{n}^2}
\label{comp}
\end{equation}
takes place, where $E_0$ is the ground state energy. In particular, vanishing  ${\cal{K}}^{-1}$ usually signals phase separated states. As also shown in Fig. \ref{fig1} and in analogy with \cite{anfossi2009}, here the single fermions move in a non-homogeneous background of empty and doubly occupied sites, forming phase separated droplets of nanoscale size. It is worthwhile to remark that the presence of such inhomogeneous superconducting state can be predicted by considering the integrable limit $\delta J=0$ $X/J=1$. Here, indeed, the nanosize structures descend from the two low-energy bands in (\ref{spec}) which acquire different bandwidth as soon as $X/J\neq 1$. As a consequence, it becomes convenient to fill more levels in one band than in the other and these two effectively different filling fix the scale of the microscopic modulation as shown in Fig. \ref{fig1}. 

\begin{figure}
\includegraphics[width=\columnwidth]{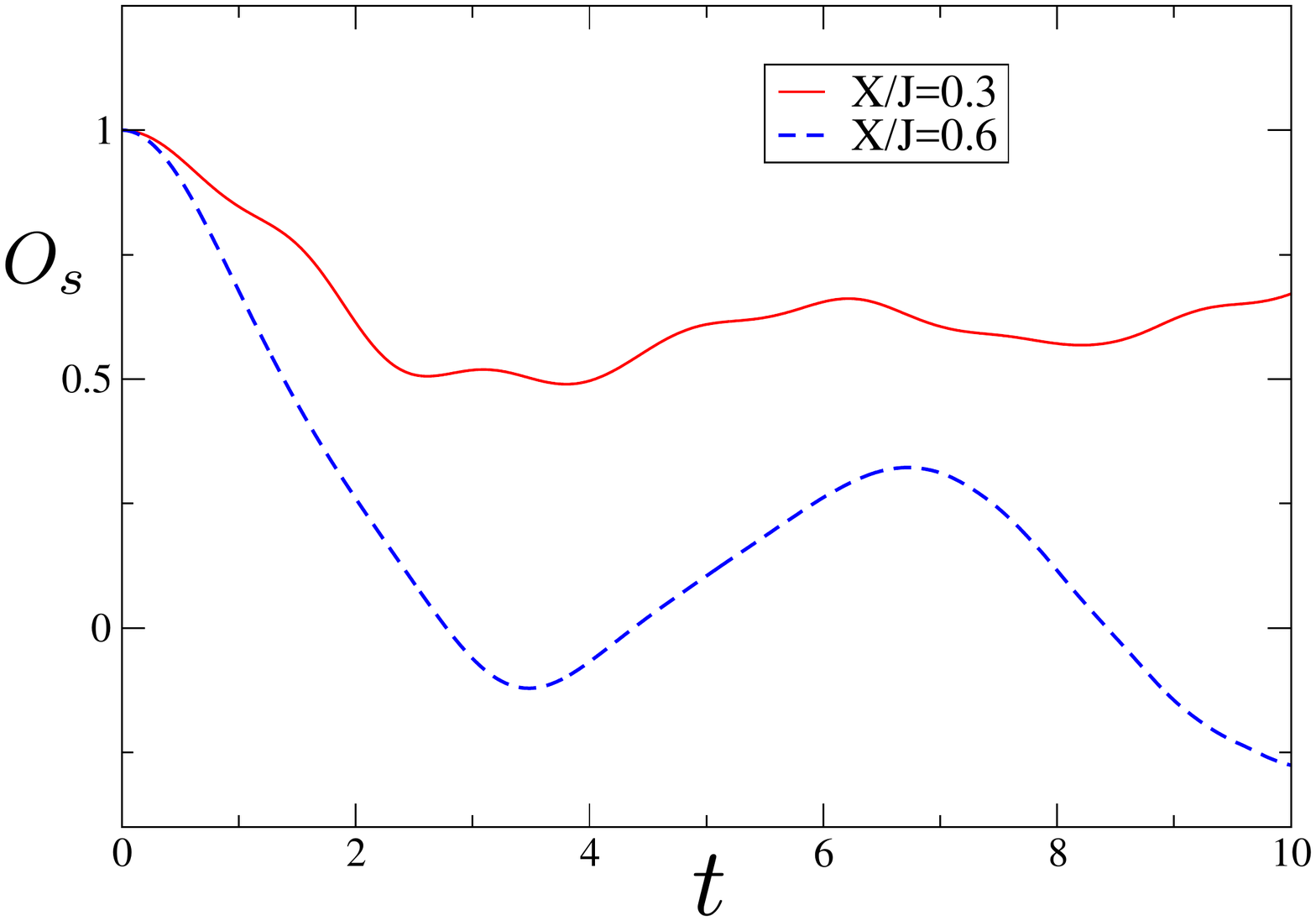}
\caption{(Color online). Time evolution of the string correlation function eq. (\ref{string}). The state at time $t=0$ is obtained by considering eq. (\ref{ssh}) at $X=0$ and $\delta J/J=0.4$. We evolve such state with the same model eq. (\ref{ssh}) by fixing $\delta J/J=0.4$ and different values of $X/J$. The time is in unit of $J^{-1}$ and we consider $N=2N_{\uparrow}=2N_{\downarrow}=L=32$.}
\label{fig3}
\end{figure}

\paragraph{\textbf{Sudden Quench dynamics}.}
As SPT phases are supposed to be unstable with respect to perturbations that break space-group symmetries, here we further underline the intriguing features of the derived second-order topological insulator by performing a quench dynamics procedure. In particular, at the initial time $t=0$ we calculate the ground state of eq. (\ref{ssh}) with $\delta J>0$ and $X=0$. As already discussed, the latter is known to be a first-order topological insulator protected by chiral symmetry. We then let this state evolve with Hamiltonian eq. (\ref{ssh}) with different values of $X\neq 0$ which, as already pointed out, break the chiral symmetry. Our time-dependent DMRG calculations \cite{white2004} in Fig. \ref{fig3}, show that the topological string order parameter in eq. (\ref{string}) evolves towards a finite value when the ratio $X/J$ supports the presence of the topological crystalline insulator derived in the previous paragraph. This result enforces even further our outcomes and, in particular, it shows that although the chiral symmetry is broken, the inversion symmetry results to be crucial to stabilize the underlying second-order topological phase. Indeed, when $X/J$ becomes larger and the crystalline order cannot be maintained at the ground state level, $O_s$ evolves towards a vanishing value therefore signaling the absence of topological order.   

\paragraph{\textbf{Experimental implementation and detection}.}
 As experiments in solid state devices have, at the moment, explored non-interacting higher order topological phases, here we discuss how ultracold atomic systems have the potentiality to explore interacting one-dimensional second order topological phases. More precisely, our model and results can be implemented and tested with the ongoing experimental platforms based on ultracold mixtures of fermionic atoms in optical lattice. Here, Floquet schemes in mixtures of ultracold fermions have indeed already allowed to engineer tunable correlated hopping process \cite{messner2018,gorg2018,Gorg2019}. At the same time, an effective lattice dimerization can be realized by means of either a super-optical lattice \cite{Atala2013,walter2022} or optical tweezers arrays \cite{Spar2022}. Finally, quantum gas microscopes have made possible to perform precise measurements of topological string correlation functions \cite{Endres2011,Hilker484,sompet2022} to detect the symmetry protected topological phase. It appears thus natural to state that an efficient and accurate quantum simulation of our results can be realized and the interacting second-order topological insulator investigated.

\paragraph{\textbf{Conclusions}.}
In this paper we shed new light on the fascinating topic of higher order topological states of matter in one-dimensional interacting quantum systems. In particular, we have introduced and studied a model where a specifically designed interaction can give rise to a second-order insulating phase. 
As we have clearly shown, even in presence of interaction described by correlated hopping processes, the combination of inversion and time reversal symmetries can indeed protect the topological nature of the insulating states characterized by long-range of non-local string correlation function and even degeneracy of the entanglement spectrum. Relevantly, we have also  discussed how our results can be tested with the last generation of atomic quantum simulators by using fermionic mixtures in optical lattice. In our knowledge, our work contains the first proposal for the quantum simulation of an interacting second-order one-dimensional topological phase with gapped edge states. As a natural perspective, we believe that the investigation of the correlated hopping interaction in two-dimensional HOTIs and fragile topological insulators might reveal further relevant features of topological crystalline phases. 

\begin{acknowledgments}

{\it \textbf{Acknowledgments:}} We thank A. Dauphin, F. Dolcini, S. Juli\`a-Farr\'e, L. Rossi, K. Vieban and B. Wieder for discussions. Work at ICFO is supported by ERC AdG NOQIA; Ministerio de Ciencia y Innovation Agencia Estatal de Investigaciones (PGC2018-097027-B-I00/10.13039/501100011033,  CEX2019-000910-S/10.13039/501100011033, Plan National FIDEUA PID2019-106901GB-I00, FPI, QUANTERA MAQS PCI2019-111828-2, QUANTERA DYNAMITE PCI2022-132919,  Proyectos de I+D+I “Retos Colaboración” QUSPIN RTC2019-007196-7); European Union NextGenerationEU (PRTR);  Fundació Cellex; Fundació Mir-Puig; Generalitat de Catalunya (European Social Fund FEDER and CERCA program (AGAUR Grant No. 2017 SGR 134, QuantumCAT \ U16-011424, co-funded by ERDF Operational Program of Catalonia 2014-2020); Barcelona Supercomputing Center MareNostrum (FI-2022-1-0042); EU Horizon 2020 FET-OPEN OPTOlogic (Grant No 899794); National Science Centre, Poland (Symfonia Grant No. 2016/20/W/ST4/00314); European Union’s Horizon 2020 research and innovation programme under the Marie-Skłodowska-Curie grant agreement No 101029393 (STREDCH) and No 847648  (“La Caixa” Junior Leaders fellowships ID100010434: LCF/BQ/PI19/11690013, LCF/BQ/PI20/11760031,  LCF/BQ/PR20/11770012, LCF/BQ/PR21/11840013. L.B. acknowledges Politecnico di Torino for the starting package grant number 54$\_$RSG21BL01 and ICFO where this project has been conceived. D.G.-C. is supported by the Simons Collaboration on Ultra-Quantum Matter, which is a grant from the Simons Foundation (651440, P.Z.)   
\end{acknowledgments}

\bibliographystyle{apsrev4-2}

\bibliography{references.bib}

\end{document}